\documentclass{article}

\usepackage{geometry}
\usepackage{amsmath}
\usepackage{calc}
\usepackage{graphicx}
\usepackage{subfigure}
\usepackage{units}

\usepackage{particles}
\usepackage{numinsec}
\usepackage{pptitle}

\title{Resonant and Non--Resonant Effects in Photon--Technipion
  Production at Lepton Colliders} 
\author{Kenneth Lane\thanks{lane@bu.edu}$^{, *, \dagger}$, 
  Kevin R. Lynch\thanks{krlynch@bu.edu}$^{, \dagger}$,
  Stephen Mrenna\thanks{mrenna@fnal.gov}$^{, *}$,\\
  and Elizabeth H. Simmons\thanks{simmons@bu.edu}$^{, \dagger, \ddagger}$\\ \\
  $^*$ Fermi National Accelerator Laboratory\\
  Batavia, IL 60510\\ \\
  $^\dagger$ Department of Physics, Boston University\\ 590 Commonwealth
  Avenue, Boston MA 02215\\ \\
  $^\ddagger$ Physics Department, Harvard University\\ Cambridge, MA 02138}
\date{March 26, 2002}
\preprintwidth{FERMILAB-PUB-02/039-T}
\preprints{BUHEP-02-14\\FERMILAB-PUB-02/039-T\\HUTP-01/A072\\hep-ph/xxyyzzz}

\newcommand{\techniV}{\particle{V_T}}
\newcommand{\technipi}{\particle{\pion_T}}
\newcommand{\technipiprime}{\particle{\pion_T^\prime}}
\newcommand{\technirho}{\particle{\rho_T}}
\newcommand{\techniomega}{\particle{\omega_T}}
\newcommand{\technipiboth}{\particle{\Pi_T}}

\newcommand{\Ftechni}{\ensuremath{F_\mathrm{T}}}
\newcommand{\NTC}{\ensuremath{N_{\mathrm{TC}}}}
\newcommand{\SUNTC}{\group{SU}{\NTC}}

\newcommand{\anom}{\ensuremath{\mathcal{V}_{G_1 G_2 \technipiboth}}} 
\newcommand{\anomggpiboth}{\ensuremath{\mathcal{V}_{\photon \photon
      \technipiboth}}} 
\newcommand{\anomggpi}{\ensuremath{\mathcal{V}_{\photon \photon
      \technipi}}} 
\newcommand{\anomggpip}{\ensuremath{\mathcal{V}_{\photon \photon
      \technipiprime}}} 
\newcommand{\anomgZpiboth}{\ensuremath{\mathcal{V}_{\photon \Znaught
      \technipiboth}}} 
\newcommand{\anomgZpi}{\ensuremath{\mathcal{V}_{\photon \Znaught
      \technipi}}} 
\newcommand{\anomgZpip}{\ensuremath{\mathcal{V}_{\photon \Znaught
      \technipiprime}}}

\newcommand{\anomGpGpiboth}{\ensuremath{\mathcal{V}_{G' G
      \technipiboth}}}

\newcommand{\propgg}{\ensuremath{\Delta_{\photon\photon}(s)}}
\newcommand{\propZZ}{\ensuremath{\Delta_{\Znaught\Znaught}(s)}}
\newcommand{\propgZ}{\ensuremath{\Delta_{\photon\Znaught}(s)}}
\newcommand{\propZg}{\ensuremath{\Delta_{\Znaught\photon}(s)}}
\newcommand{\propgV}{\ensuremath{\Delta_{\photon V_T}(s)}}
\newcommand{\propZV}{\ensuremath{\Delta_{\Znaught V_T}(s)}}
\newcommand{\propgGp}{\ensuremath{\Delta_{\photon G'}(s)}}
\newcommand{\propZGp}{\ensuremath{\Delta_{\Znaught G'}(s)}}

\newcommand{\QU}{\ensuremath{Q_\particle{U}}}
\newcommand{\QD}{\ensuremath{Q_\particle{D}}}
\newcommand{\cchi}{\ensuremath{c_\chi}}
\newcommand{\cchip}{\ensuremath{c_{\chi'}}}

\newcommand{\Vcoupling}{\ensuremath{V_{V_T \photon \technipiboth}}}
\newcommand{\Acoupling}{\ensuremath{A_{V_T \photon \technipiboth}}}
\newcommand{\Xcoupling}{\ensuremath{X_{V_T \photon \technipiboth}}}
\newcommand{\VVGpi}{\ensuremath{V_{V_T G \technipiboth}}}
\newcommand{\AVGpi}{\ensuremath{A_{V_T G \technipiboth}}}

\newcommand{\lambdafunc}[3]{\ensuremath{\lambda({#1},{#2},{#3})}}

\newcommand{\GVpL}{\ensuremath{\mathcal{G}_{\electron
      L}^{V\photon\technipiboth}}} 
\newcommand{\GVpR}{\ensuremath{\mathcal{G}_{\electron
      R}^{V\photon\technipiboth}}} 
\newcommand{\GApL}{\ensuremath{\mathcal{G}_{\electron
      L}^{A\photon\technipiboth}}} 
\newcommand{\GApR}{\ensuremath{\mathcal{G}_{\electron
      R}^{A\photon\technipiboth}}} 
\newcommand{\GXlambda}{\ensuremath{\mathcal{G}_{\electron
      \lambda}^{X\photon\technipiboth}}}

\newcommand{\GVL}{\ensuremath{\mathcal{G}_{\electron L}^{V G
      \technipiboth}(s)}} 
\newcommand{\GVR}{\ensuremath{\mathcal{G}_{\electron R}^{V G
      \technipiboth}(s)}} 
\newcommand{\GAL}{\ensuremath{\mathcal{G}_{\electron L}^{A G
      \technipiboth}(s)}} 
\newcommand{\GAR}{\ensuremath{\mathcal{G}_{\electron R}^{A G
      \technipiboth}(s)}}
\newcommand{\GVgen}{\ensuremath{\mathcal{G}_{\electron \lambda}^{V
      G \technipiboth}(s)}} 
\newcommand{\GAgen}{\ensuremath{\mathcal{G}_{\electron \lambda}^{A
      G \technipiboth}(s)}}  

\newcommand{\missingE}{\DS{E}}

\makeatletter
\renewcommand{\@thesubfigure}{(\alph{subfigure})\space}
\renewcommand{\p@subfigure}{}
\makeatother

\newcommand{\crosssec}[1]{{\ensuremath{\mathrm{\sigma}({#1})}}}
\DeclareMathOperator{\trace}{Tr}

\begin{document}

\begin{titlepage}
\maketitle

\begin{abstract}

Lepton collider experiments can search for light technipions in final
states made striking by the presence of an energetic photon: $\epem
\to \photon\technipiboth$.  To date, searches have focused on either
production through anomalous coupling of the technipions to
electroweak gauge bosons or on production through a technivector meson
(\technirho, \techniomega) resonance.  This paper creates a combined
framework in which both contributions are included.  This will allow
stronger and more accurate limits on technipion production to be set
using existing data from LEP or future data from a higher-energy
linear collider.  We provide explicit formulas and sample calculations
(analytic and Pythia) in the framework of the Technicolor Straw Man
Model, a model that includes light technihadrons.

\end{abstract}

\end{titlepage}
\section{Introduction}
 
Modern technicolor \cite{technicolor} models
require a walking gauge coupling
\cite{Holdom:1981bg,walking}
to avoid large flavor--changing neutral currents and extra top quark
dynamics such as topcolor interactions \cite{topcolor}
to generate the large top quark mass.  To incorporate these
innovations, a large number of technifermion doublets, $N_D$, must
typically be present in the model to perform such crucial tasks as
flattening the beta function and breaking the topcolor interactions to
ordinary color
\cite{lane,Eichten:1996dx,Eichten:1998kn,Eichten:1997yq}.
This large number of doublets also suppresses the
technihadron mass scale, resulting in a small technipion decay
constant
\begin{equation}
\Ftechni \approx \frac{\unit[246]{GeV}}{\sqrt{N_D}}\ ,
\end{equation}
and very light technipions \cite{lane}.  For
example, if $N_D = 10$, $\Ftechni \approx \unit[80]{GeV}$.  With such
a low mass scale, the question of collider phenomenology becomes of
immediate interest, since the lowest lying technimeson states could be
produced directly at current or near future experiments 
\cite{tevatron,lep}.

This study discusses production of light technimeson states at lepton
colliders.  We focus on providing a complete phenomenological
description of both resonant and non-resonant technimeson production.
The framework created here should enable the LEP experiments to obtain
more comprehensive limits on light technihadrons from their final
analyses than have been extracted thus far with the more limited
methods available previously \cite{lep,Lynch:2000hi}.  We
perform several sample calculations for colliders with $\sqrt{s}$ of
up to a few hundred GeV, consistent with our interest in the LEP data.
However, our methods are also applicable to future linear
colliders\footnote{Technically, our methods can also be used for
hadron colliders, but the larger backgrounds there typically render
non-resonant production invisible.}  at higher energies.

 We use the Technicolor Straw Man Model (TCSM)
\cite{Lane:1999uk,Lane:1999uh} as a benchmark for assessing the
experimental visibility of technipion production.  The TCSM assumes
that techni--isospin is a good symmetry, and that, in analogy with
QCD, the lightest technimesons are constructed solely from the
lightest technifermion weak doublet,
$(\particle{T_U},\particle{T_D})$, which transform as $\SUthree_C$
singlets and $\SUNTC$ fundamentals.  The members of the doublet are
assigned electric charges \QU\ and \QD\ respectively.  This flavor and
gauge structure gives rise to the same type of spectrum as two--flavor
QCD: namely, an isotriplet and isosinglet of pseudoscalar,
pseudo-goldstone modes, the $\technipi^{0,\pm}$ and
$\technipiprime^0$, and an isotriplet and isosinglet of vector modes,
the $\technirho^{0,\pm}$ and $\techniomega^0$.  The electric charge
assignments of the mesons require that $\QU - \QD = 1$.  Since we
assume that techni--isospin symmetry is a good symmetry, the
technipions should be nearly degenerate in mass, as should the
technivector modes.  When both the \technipi\ and \technipiprime\ are
possible final states for a given process, we will refer to both
collectively by the notation \technipiboth.

\begin{figure}
\begin{center}
\subfigure[Anomaly model]{
\includegraphics[width=(\textwidth-1in)/2]{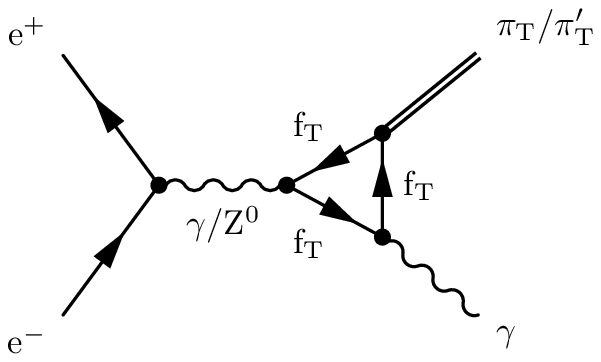}
\label{fig:anomaly}} \qquad\qquad
\subfigure[\techniV-dominance model]{
\includegraphics[width=(\textwidth-1in)/2]{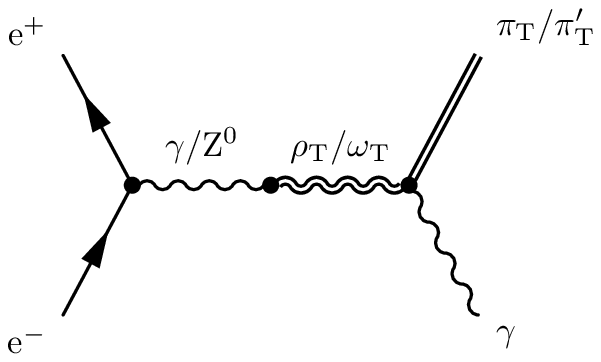}
\label{fig:kinetic}} 
\caption{\ref{fig:anomaly} The anomaly--mediated production mechanism
  of a $\photon\technipiboth$ final state.  \ref{fig:kinetic} The
  \techniV--dominated production mechanism to lowest order in $\alpha$.}
\label{fig:diagrams}
\end{center}
\end{figure}

Calculation of matrix elements involving the technihadron bound states
at energies below the technicolor scale, $\Lambda_{\mathrm{TC}}$, in
the full non-abelian technicolor model requires use of low energy
phenomenological models.  In the recent past, two different types of
descriptions have been widely used for fermion--antifermion
annihilation to a technipion plus electroweak gauge boson.  In these the
initial--state fermions couple with standard weak couplings to the
appropriate electroweak gauge bosons in the $s$--channel.  The
descriptions differ in how they handle the weak gauge boson transition
to the final state, and can be divided into
\begin{enumerate}
\item Anomaly--Mediated Approach: the gauge boson couples to the
  \technipiboth\ in the final state through a technifermion
  ($\particle{f_T}$) triangle anomaly
  \cite{Manohar:1990eg,Randall:1992gp} (Figure~\ref{fig:anomaly}), and
\item Technivector (\techniV) Dominated Approach: the gauge boson
undergoes a kinetic mixing (that is, a term proportional to $s$ in the
inverse propagator matrix) into a \technirho\ or \techniomega, which
then decays directly into the \technipiboth\ in the final state
\cite{Eichten:1996dx,Lane:1999uk,Lane:1999uh}
(Figure~\ref{fig:kinetic}).
\end{enumerate}
Both schemes have direct analogues in Standard Model QCD calculations.
Our goal is to synthesize these approaches within the TCSM to
eliminate the shortcomings of each.

In Section~\ref{sec:approaches}, we review the details of both the
anomaly--mediated and \techniV--dominated approaches to the TCSM and
indicate the limitations of their individual descriptions of
technimeson production at lepton colliders.  Our calculations focus on
the process $\epem \to \photon\technipiboth$ because kinematic and
phase space considerations should give it a larger cross--section than
processes involving final--state weak bosons.  In
Section~\ref{sec:combining} we discuss how to combine the
strengths of both approaches within the TCSM framework.  In
Section~\ref{sec:pit-photon} we compare analytic cross section
predictions for $\epem \to \photon\technipiboth$ in all three
approaches.  In Section~\ref{sec:pythia} we discuss the predictions
for the mass recoiling against the photon in the process $\epem \to
\photon\technipiboth$ within the combined framework as implemented in
Pythia.

\section{Phenomenological Approaches}
\label{sec:approaches}

\subsection{Anomaly Mediation}
\label{sec:anomaly}

In the anomaly--mediated schemes, we assume that the lowest lying
observable states, the \technipi\ and \technipiprime, are
pseudo--Nambu--Goldstone modes of the \SUNTC\ technicolor theory, in
direct analogy to the QCD pion.  The coupling of these \technipiboth\ 
goldstone modes to a pair, $G_1$ and $G_2$, of electroweak gauge
bosons is given by \cite{dimopoulos,Holdom:1981bg}
\begin{equation}
\mathcal{M} (G_1(q) \to G_2(p_1) \technipiboth(p_2)) = \NTC \anom \frac{g_1 
  g_2}{8 \pi^2 \Ftechni} \varepsilon^{\mu\nu\lambda\rho}
    \epsilon_\mu(q) \epsilon^*_\nu(p_1) q_\lambda p_{1\rho}
\label{eq:anomaly-coupling}
\end{equation}
where \NTC\ is the number of technicolors, the $g_i$ are the couplings
of the gauge groups, $q$ and $p_1$ are the momenta and the
$\epsilon_i$ the polarizations of the gauge bosons.\footnote{We choose
  our notation to agree with \cite{Lane:1999uk}.}  The triangle
anomaly factor, \anom, is given by
\cite{dimopoulos,Holdom:1981bg}  
\begin{equation}
  \anom = \trace\left[T^a \left(\left\{T_1, T_2\right\}_L +
    \left\{ T_1, T_2\right\}_R\right)\right]\ .  
\end{equation}
Here $T_i$ is the generator associated with the gauge boson $G_i$, and
$T^a$ is the generator of the axial current associated with the
technipion
\begin{equation}
j^{\mu a}_5 = \DB{\psi} \DG{\mu} \DGL{5} T^a \psi\ ;
\end{equation} 
in this convention, the generators are normalized such that
$\trace{\left(T^a T^b\right)} = \frac{1}{2} \delta^{ab}$ .

Using these expressions, we can calculate the cross section for
$\epem \to \photon\technipiboth$ in an anomaly mediated framework, 
obtaining \cite{Randall:1992gp} 
\begin{multline}
  \crosssec{\epem \to \photon\technipiboth} = \frac{\alphaem^3
    \NTC^2}{192 \pi^2 \Ftechni^2} \left(s - M_\technipiboth^2\right)^{3/2}\\
  \times \left[
    \left|\left( \anomggpiboth \propgg -
        \frac{2 \zeta_{\electron L} \anomgZpiboth}{\sin 2\thetaW}
        \propZZ \right)\right|^2 + 
    \left|\left( \anomggpiboth \propgg -
        \frac{2 \zeta_{\electron R}\anomgZpiboth}{\sin 2\thetaW} \propZZ
      \right)\right|^2 
  \right]\ ,
\label{eqn:anomalyxsec}
\end{multline}
where $\propgg^{-1} = s$, $\propZZ^{-1} = s -M_\Znaught^2 + i
\Gamma_\Znaught M_\Znaught$, and $\zeta_{\electron \lambda} =
T^3_{\electron_\lambda} + \sin^2\thetaW$ for chiralities
$\lambda=L,R$. For the TCSM, the anomaly factors involving the
\technipi\ and \technipiprime\ are given by
\cite{Lynch:2000hi,Lane:1999uk}  
\begin{align}
\anomggpi &= 2 \left( \QU + \QD \right) \cchi & \anomgZpi &=
\frac{\left( \QU + \QD \right) \left(1 - 4 \sin^2\thetaW \right)
  \cchi}{\sin 2\thetaW}\\
\anomggpip &= 2 \left( \QU^2 + \QD^2 \right) \cchip & \anomgZpip &=
\frac{\left(1 - 4 \sin^2\thetaW \left( \QU^2 + \QD^2 \right)  \right)
  \cchip}{\sin 2\thetaW}\ .
\end{align}
A more detailed discussion of this and other processes in the anomaly
framework can be found in
\cite{Randall:1992gp,Rupak:1995kg,Lynch:2000hi}.  Using this
framework, limits on various TC models have been extracted
from published LEP data on final states with photons, large missing
energy, jet pairs, or $b\bar{b}$ pairs in \cite{Lynch:2000hi}.
Production of technipions in the anomaly framework at future \epem\
colliders has been discussed in~\cite{Casalbuoni}

The anomaly mediated description has the dual strengths of conceptual
clarity and relative ease of calculation.  It does, however, have a
flaw which would not be present in a complete technicolor model and
and which prevents it from being an appropriate description in all
kinematic regimes.  Since this scheme does not take into account the
heavier technimeson bound states of the technifermions (the states
equivalent to the QCD $\rho$ and $\omega$, among others), it can only
provide a valid description of technicolor physics in kinematic
regions well below the propagator poles of the lightest technivector
meson \cite{Lane:1999uk}.

\subsection{Technivector Meson Dominance in the TCSM}
\label{sec:kinetic}

To describe the kinematic regime near the technivector poles, an
alternative phenomenological approach is needed; typically this takes
the form of the \techniV--dominance scheme introduced above.  For the
TCSM, a framework has been developed by Lane
\cite{Lane:1999uk,Lane:1999uh}.  Conceptually, the collider
experiments generate electroweak gauge bosons via the direct couplings
of standard model particles to the electroweak gauge fields.  The
electroweak gauge bosons then convert into technivector mesons through
mixing terms in the vector propagator matrix (for illustration, we
display the inverse of the neutral propagator matrix here)
\begin{equation}
\Delta_0^{-1}(s) = \left(
\begin{array}{cccc}
s & 0 & s f_{\photon\technirho} & s f_{\photon\techniomega}\\
0 & s - \mathcal{M}_\Znaught^2 & s f_{\Znaught\technirho} & s
f_{\Znaught\techniomega}\\ 
s f_{\photon\technirho} & s f_{\Znaught\technirho} & s -
\mathcal{M}_\technirho^2 & 0 \\
s f_{\photon\techniomega} & s f_{\Znaught\techniomega} & 0 & s -
\mathcal{M}_\techniomega^2 
\end{array}
\right)
\begin{array}{c}\\ \\ \\ \ , \end{array}
\end{equation}
where the masses, $\mathcal{M}_V = M_V^2 - i \sqrt{s} \Gamma_V$
include $s$--dependent width effects.  The mixing factors are
$f_{\photon\technirho} = \xi$, $f_{\photon\techniomega} = \xi (\QU +
\QD)$, $f_{\Znaught\technirho} = \xi \cot 2 \theta_W$, and
$f_{\Znaught\techniomega} = -\xi (\QU + \QD)\tan\theta_W$ where $\xi^2
= \alpha_{em}/\alpha_{\technirho}$ \cite{Lane:1999uk,Lane:1999uh}.  These
vector technimesons decay into the lighter spinless technimesons,
electroweak bosons, and fermion--antifermion pairs.

The TCSM was developed to describe technihadron production at
high--energy hadron colliders for which the convoluted parton
distributions sweep over the $\technirho/\techniomega$ resonance
poles. In its original form, it did not properly include contributions
that are far below the poles \cite{Lane:1999uk}. However, at an \epem\ 
collider such as LEP (or a future linear collider), the
machine's operating energy $\sqrt{s}$ may be well away from the resonance.
For those cases, it is necessary to include
off--resonance contributions. At the very least, this may allow more
stringent limits on technihadron masses and couplings to be derived
from searches in $\epem$ colliders.

The coupling of the initial state electrons to the gauge boson is
unchanged from the Standard Model.  The couplings of the
technivectors, \technirho\ and \techniomega, to the final state
technipion and photon are given by the TCSM matrix element
\cite{Lane:1999uk}
\begin{equation}
  \begin{split}
    \mathcal{M} \left( V_T(q) \to \photon(p_1) \technipiboth(p_2)\right) = &
    \frac{e \Vcoupling}{M_V} \varepsilon^{\mu\nu\lambda\rho}
    \epsilon_\mu(q) \epsilon^*_\nu(p_1) q_\lambda p_{1\rho} + \\
    & \frac{e \Acoupling}{M_A} \left( \epsilon(q)\cdot
      \epsilon^*(p_1) q\cdot p_1 - \epsilon(q)\cdot p_1
      \epsilon^*(p_1)\cdot q \right)\ , 
  \end{split}
  \label{eq:kinetic-coupling}
\end{equation}
where the first (second) term is the vector (axial) contribution, and
$M_V$ and $M_A$ are dynamical mass parameters of the same order that
set the strengths of these terms (for simplicity we set them equal
below).  The relevant axial couplings ($\Acoupling$) are zero; the
relevant vector ($\Vcoupling$) couplings are \cite{Lane:1999uk} 
\begin{align}
V_{\technirho\photon\technipi} 
  &= 2 \left( \QU + \QD \right) \cchi & 
V_{\technirho\photon\technipiprime}  &= \cchip\\
V_{\techniomega\photon\technipi} &= \cchi &
V_{\techniomega\photon\technipiprime} 
&= 2 \left( \QU + \QD \right) \cchip  .
\end{align}
A list of analogous couplings for other gauge bosons and $\techniV$ in
the TCSM is given in \cite{Lane:1999uk}.

The cross section for
$\epem \to \photon\technipiboth$ is given by \cite{Lane:1999uk}
\begin{equation}
  \crosssec{\epem \to \photon\technipiboth} = \frac{\pi \alphaem^2}{108
    M_V^2} \frac{\left(s - M_\technipiboth^2\right)^4}{s^2} 
  \left[ 
    \left|\GVpL\right|^2 + \left|\GVpR\right|^2 +
    \left|\GApL\right|^2 + \left|\GApR\right|^2 
  \right]\ .
\label{eqn:VTxsec}
\end{equation}
The \GXlambda\ are given by
\begin{equation}
\GXlambda = \sum_{V_T = \technirho,\techniomega} \Xcoupling
\mathcal{F}_{\electron\lambda}^{V_T}(s)\ ,
\end{equation} 
where the \Xcoupling\ are the vector and axial couplings of the vector
technimesons to the technipion and photon, and 
\begin{equation}
\mathcal{F}_{\electron\lambda}^{V_T}(s) = e \Delta_{\photon V_T}(s) + \frac{2
  \zeta_{\electron\lambda}}{\sin 2\thetaW} \Delta_{\Znaught V_T}(s)
\end{equation}
includes the coupling of the initial state electrons to the gauge
bosons and the propagator elements that mix the vector bosons with the
technivector mesons.  A more detailed discussion of this and other
processes in the \techniV--dominance approach to the TCSM can be found
in \cite{Lane:1999uk,Lane:1999uh}.  The DELPHI and OPAL experiments at
LEP \cite{lep} used \techniV--dominance to obtain limits on $\epem \to
\technirho, \techniomega \to \photon\technipiboth$ and related
processes in the TCSM.

\subsection{Combining Both Schemes}
\label{sec:combining}

The center of mass energies of LEP and proposed future linear
colliders are comparable to the expected masses of the lightest
technihadrons in low--scale technicolor: a few hundred GeV.  Hadron
collider experiments are sensitive only to resonant technivector
contributions, and therefore need consider only contributions from the
poles.  In contrast, lepton collider experiments have a broader
sensitivity and may well be operating off the poles --- especially
when their location is unknown. For an \epem\ collider operating
slightly below the poles, it is especially important to understand how
the resonant and non--resonant contributions are combined.

Schematically, we would like to define a matrix element that
interpolates between the anomaly mediated matrix element at the
\technipiboth\ production threshold and the \techniV-dominated matrix
element in the region of the technivector poles, that is
\begin{equation*}
\mathcal{M}_{\text{combined}} = \mathcal{M}_{\text{anomaly}}
\left[f(s)\right]^n + \mathcal{M}_{\techniV} \left[1-f(s)\right]^n\ ,
\end{equation*}
where the interpolating function $f(s)$ has the limits $f(s \to 0) \to
1$ and $f(s \to M_{\technirho,\techniomega}) \to 0$.  Numerically, we
find that either the anomaly--mediated or the \techniV--mediated
matrix element completely dominates the cross section, except in a
relatively narrow region approximately midway between threshold and
the first technivector pole, where they are of approximately equal
magnitude (see Figure~\ref{fig:xsecs}).  Because of this behavior, we
gain little by implementing such a complicated scheme rather than
simply taking $n\to0$ in the above interpolation, that is, simply
adding the relevant matrix elements everywhere.  This gives us the
correct limits, up to numerically irrelevant errors.

Adding the matrix elements has several virtues: it reproduces the
correct cross-section both well below and in the region of the
technivector meson resonances, and it is simple to implement.  In
addition, as will be shown shortly, the combined cross-sections still
respect unitarity bounds in the energy range of experimental interest.
At much higher energies, our description will break down because
additional resonances and continuum technifermion production will
emerge, but that is not relevant to our purposes.

Since the matrix element in Equation~\ref{eq:anomaly-coupling} for the
anomaly--mediated coupling and the vectorial component in
Equation~\ref{eq:kinetic-coupling} of the matrix element in the
$\techniV$-dominated scheme have the same Lorentz structure, we add
them.  From the combined matrix elements, we obtain the cross section
for $\epem \to G \technipiboth$, where $G$ is a photon or a
transversely polarized \Znaught:
\begin{multline}
    \crosssec{\epem \to G \technipiboth} = \\ 
    \frac{\pi \alphaem^2}{12 s}
    \lambdafunc{s}{M_G^2}{M_\technipiboth^2}^{3/2} \left(
      \left|\GVL\right|^2 + \left|\GVR\right|^2 + \left|\GAL\right|^2 +
      \left|\GAR\right|^2 
    \right) \\
    + \frac{\pi \alphaem^2 M_G^2}{2}
    \lambdafunc{s}{M_G^2}{M_\technipiboth^2}^{1/2} \left(
      \left|\GAL\right|^2 + \left|\GAR\right|^2 \right)\ .
\end{multline}
Here $\lambdafunc{a}{b}{c} = a^2 +b^2 +c^2 -2ab -2ac -2bc$ and $M_G$
is the mass of the final state gauge boson.  The vectorial couplings
for a given fermion helicity $\lambda$, including both \techniV\ and
anomaly terms, is given by
\begin{multline}
  \GVgen = \sum_{V_T = \technirho,\techniomega} \frac{\VVGpi}{M_V}
  \left(Q_\elec \propgV + \frac{2 \zeta_{\elec
        \lambda}}{\sin 2\thetaW} \propZV \right) \\
  + \frac{e \NTC}{8 \pi^2 \Ftechni} \sum_{G' = \photon,\Znaught}
    \anomGpGpiboth \left(Q_\elec \propgGp + \frac{2
        \zeta_{\elec \lambda}}{\sin 2\thetaW} \propZGp \right)\ ,
\end{multline}
where in contrast to Equation~\ref{eqn:anomalyxsec}, the anomaly
contribution now includes off--diagonal mixing terms in the
propagator, \propZg\ and \propgZ, that are induced by the
presence of the \technirho\ and \techniomega\ in the vector spectrum.
The axial couplings are given by
\begin{equation}
  \GAgen = \sum_{V_T = \technirho,\techniomega} \frac{\AVGpi}{M_V}
  \left(Q_\elec \propgV + \frac{2 \zeta_{\elec
        \lambda}}{\sin 2\thetaW} \propZV \right)\ .
\end{equation}
Once again, the $\Znaught\epem$ coupling is $\zeta_{\elec \lambda} =
T^3_{\elec_\lambda} - Q_\elec \sin^2\thetaW$.

This method of combining the anomaly and \techniV\ contributions
applies more generally to fermion--antifermion annihilation into
technipion plus transverse weak gauge boson at lepton and hadron
colliders. The set of all such differential cross sections including
anomaly and \techniV\ terms and a tabulation of the various anomaly
factors in the TCSM will appear in an updated version of
\cite{Lane:1999uk}.

\section{$\epem \to \photon\technipiboth$}
\label{sec:pit-photon}

As an example of our results, we study in this section the process
$\epem \to \photon\technipiboth$, both analytically and by means of
Pythia simulations.  We remind the reader that \technipiboth\ refers
to both the \technipi\ and \technipiprime.  They can not be
distinguished experimentally unless the \technipi\ and \technipiprime\ 
have significantly different masses and/or decay modes.  Note,
however, that interference between production of \technipi\ and
\technipiprime\ decaying to the same final state will not generally be
significant because the \technipiboth\ states are extremely narrow
\cite{Lane:1999uh,Lane:1999uk}.  Only for $| M_\technipi -
M_\technipiprime | \leq \Gamma_\technipi, \Gamma_\technipiprime$ would
this be a concern.  To represent the general expectations in the TCSM,
we take $M_\technipi = M_\technipiprime$ throughout this section and
in Pythia, but do not include interference between the \technipiboth .

\subsection{Analytical Results}
\label{sec:analytical}

\begin{figure}
\begin{center}
\subfigure[{\unit[50]{GeV}} \technipiboth]{
  \includegraphics[width=(\textwidth-1in)/2]{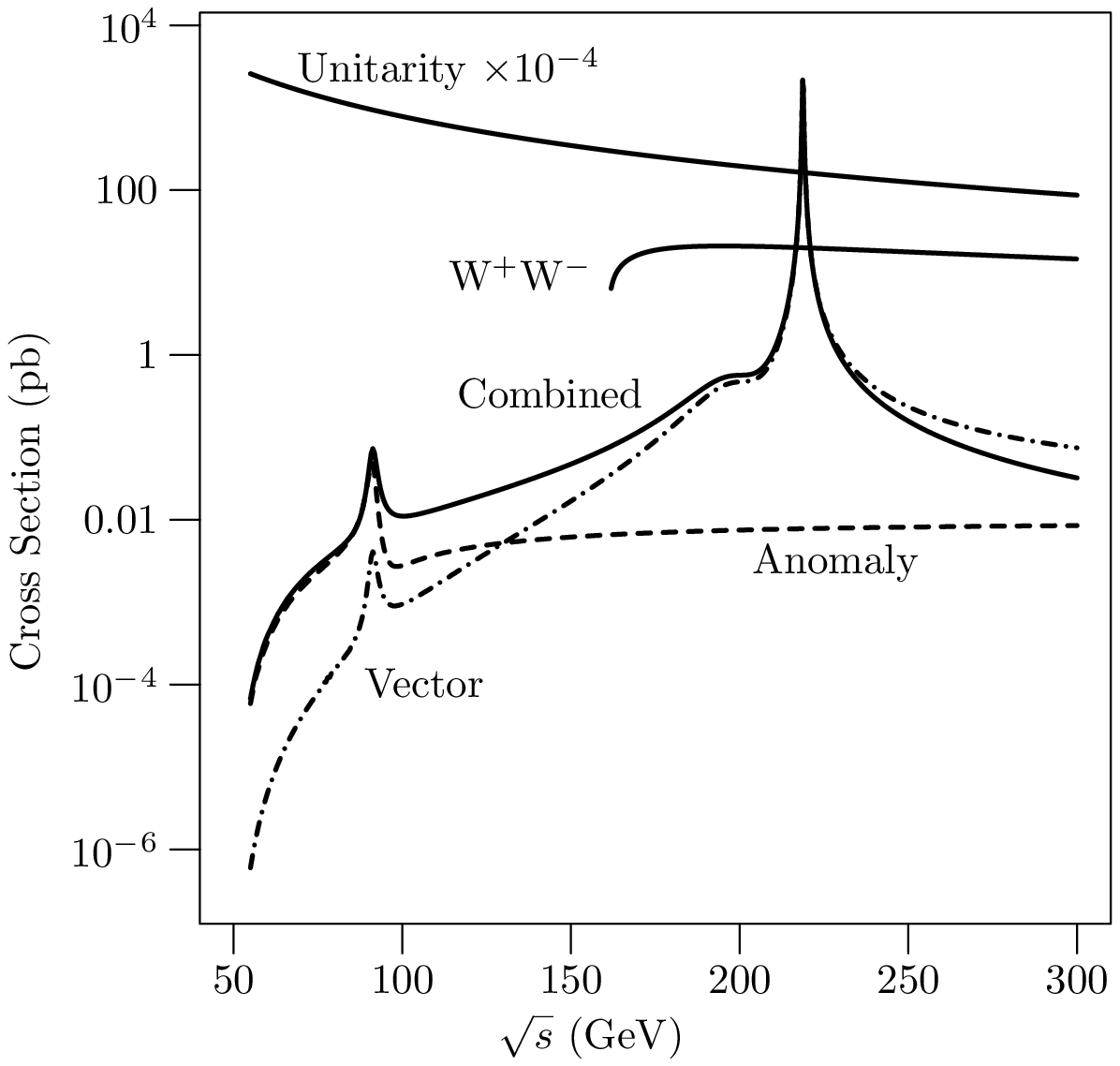}
  \label{fig:light}} \qquad\qquad
\subfigure[{\unit[110]{GeV}} \technipiboth]{
  \includegraphics[width=(\textwidth-1in)/2]{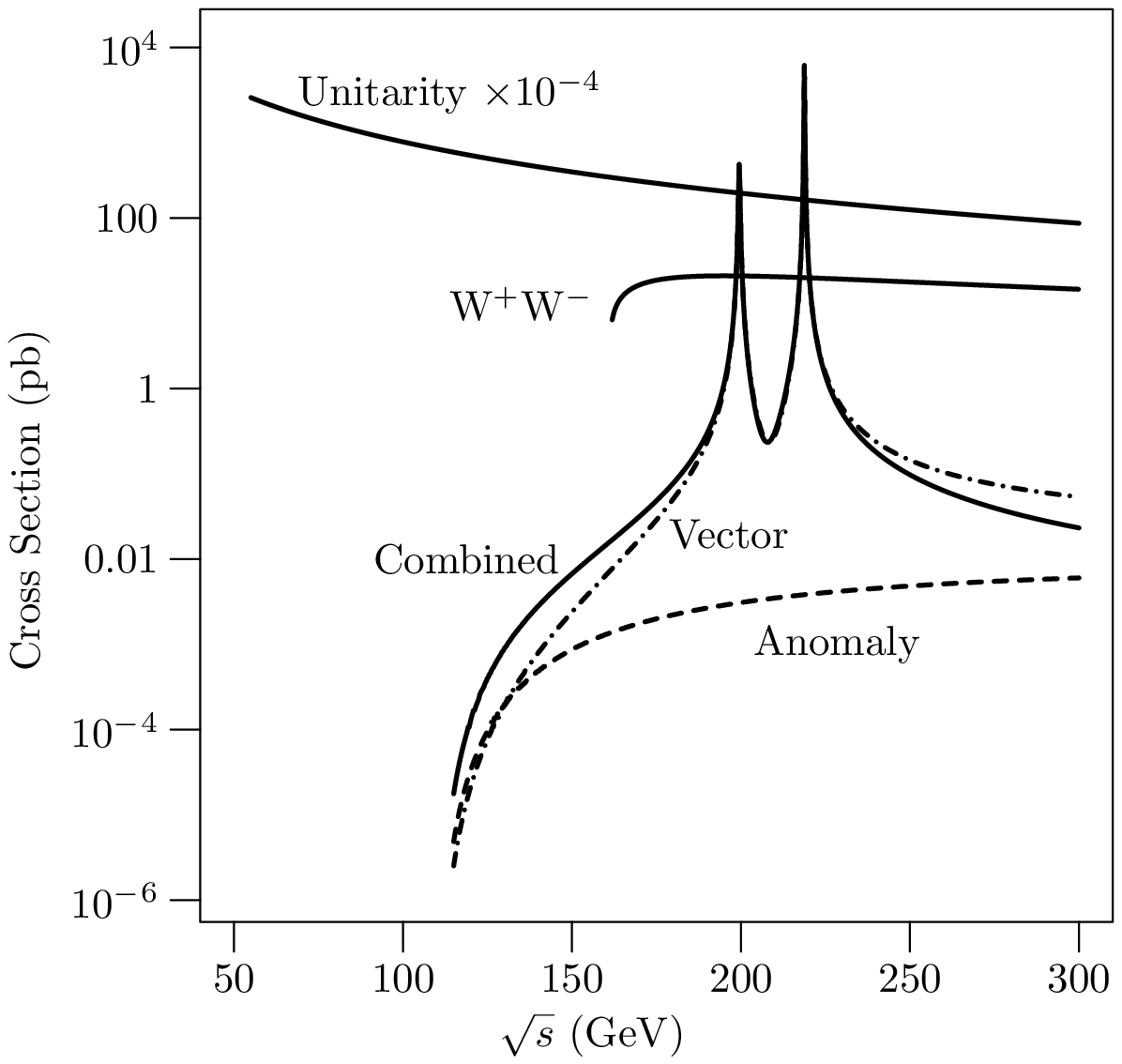}
  \label{fig:heavy}} 
\caption{These plots display $\epem \to \photon\technipiboth$ cross
sections for two \technipi\ masses (\unit[50]{GeV} on the left, and
\unit[110]{GeV} on the right), for fixed \technirho\ and \techniomega\
masses (\unit[200]{GeV} and \unit[220]{GeV} respectively), as a
function of the collider center of mass energy, $\sqrt{s}$.  Displayed
are the total cross sections for the anomaly scheme (the dashed
curves), the \techniV--dominance scheme (the dash--dotted curves), and
a scheme including both contributions (the solid curve).  In figure
\ref{fig:light}, we can see that the anomaly scheme provides the
dominant contribution at energies well below the resonances, while the
\techniV's dominate in the region of the poles.  The transition region
is quite narrow.  In figure \ref{fig:heavy}, the cross-section probes
only the region near the resonances.  For comparison purposes we also
show the unitarity limits for a process with a vector intermediary
(top solid curve) and the tree--level standard model $\epem \to
\Wplus\Wminus$ cross section (central solid curve).  }
\label{fig:xsecs}
\end{center}
\end{figure}

As noted before, in the region of the technivector poles, the
\techniV\ mesons dominate other contributions, while well below the
poles, the anomaly dominates.  In between, there is a transition
region where the contributions should be of the same order, and
neither can be considered in isolation. Because (for our choice of
technihadron masses) this is the region in which LEP experiments were
done, the combined amplitudes may result in better limits on
low--scale technicolor. In Figure~\ref{fig:xsecs}, we plot the cross
sections for the process $\epem \to \photon\technipiboth$ for two
\technipi\ masses (\unit[50]{GeV} in Figure~\ref{fig:light} and
\unit[110]{GeV} in Figure~\ref{fig:heavy}), with a \technirho\ mass of
\unit[200]{GeV} and a \techniomega\ mass\footnote{One usually expects
$M\technirho \approx M_\techniomega$ in the TCSM; here they are given
somewhat different values so the characteristics of the resonances can
be compared in the figures.} of \unit[220]{GeV}.  From these plots,
the low energy anomaly dominance and pole region \techniV--dominance
are clear. The \techniomega\ resonance is stronger than the
\technirho, because the \techniomega\ is narrower. We also see that
the transition region is relatively narrow.  For comparison with
typical weak scale processes, we also plot the tree level Standard
Model prediction for $\epem \to \Wplus\Wminus$ \cite{Brown:1979mq}.

Finally, we must ensure that the cross sections calculated above do
not violate unitarity in any kinematic region of interest.  For a
vector mediated interaction, both $l = 0$ and $l = 1$ partial waves
contribute to the cross section, and the upper bound on the cross
section from partial wave unitarity is given by $\sigma < 64 \pi / s$;
this unitarity limit is also plotted in Figure~\ref{fig:xsecs}.  The
total cross section is well within the unitarity limit in all
currently accessible kinematic regions.  Unitarity will be lost at
inaccessibly high energies, but well before that point the model
becomes invalid since it does not include higher mass
technihadrons or continuum technifermion production.

\subsection{Pythia Simulations}
\label{sec:pythia}

\begin{figure}
\begin{center}
\subfigure[Anomaly Included]{
\includegraphics[width=(\textwidth-1in)/2]{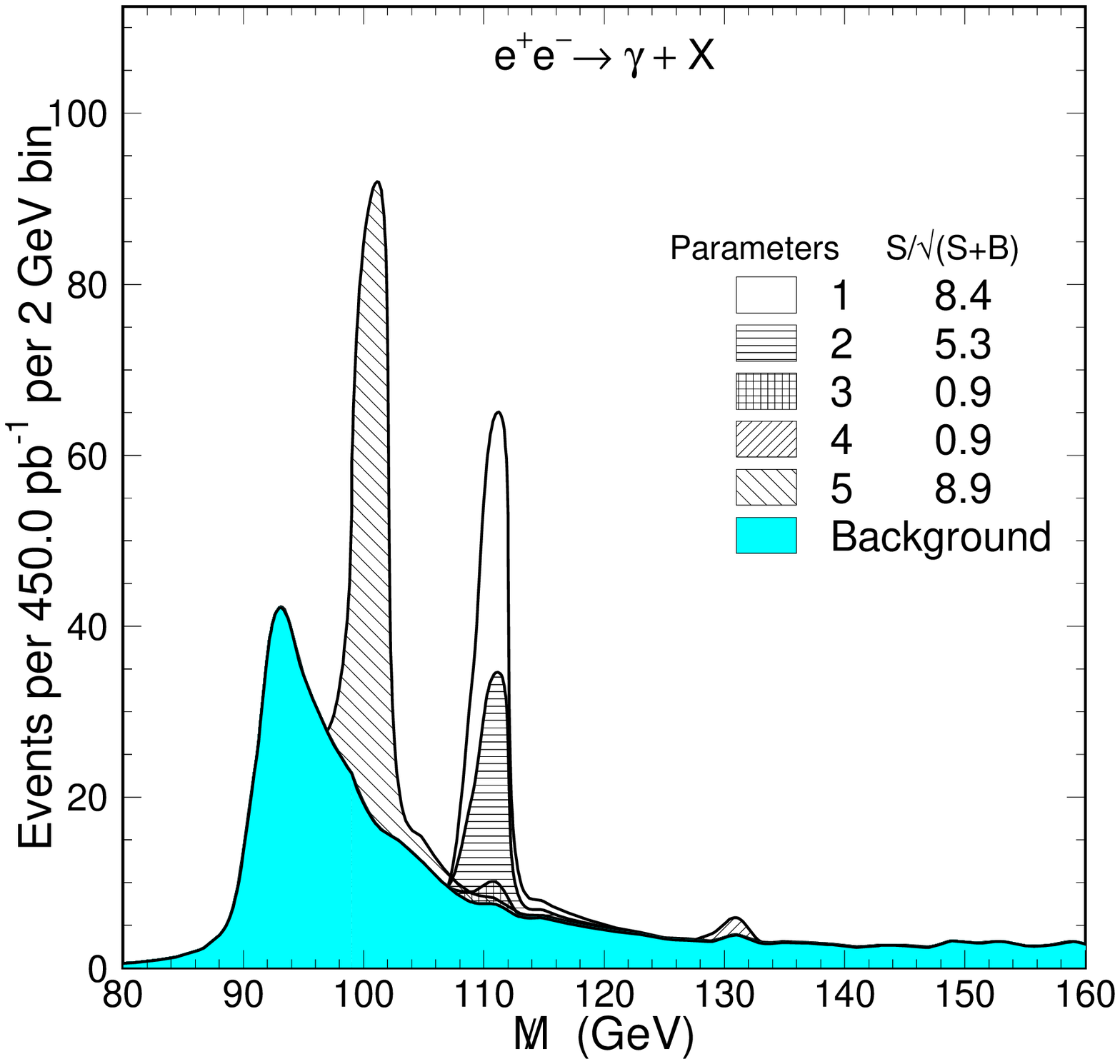}
\label{fig:pythia-a}}\qquad\qquad
\subfigure[Anomaly Excluded]{
\includegraphics[width=(\textwidth-1in)/2]{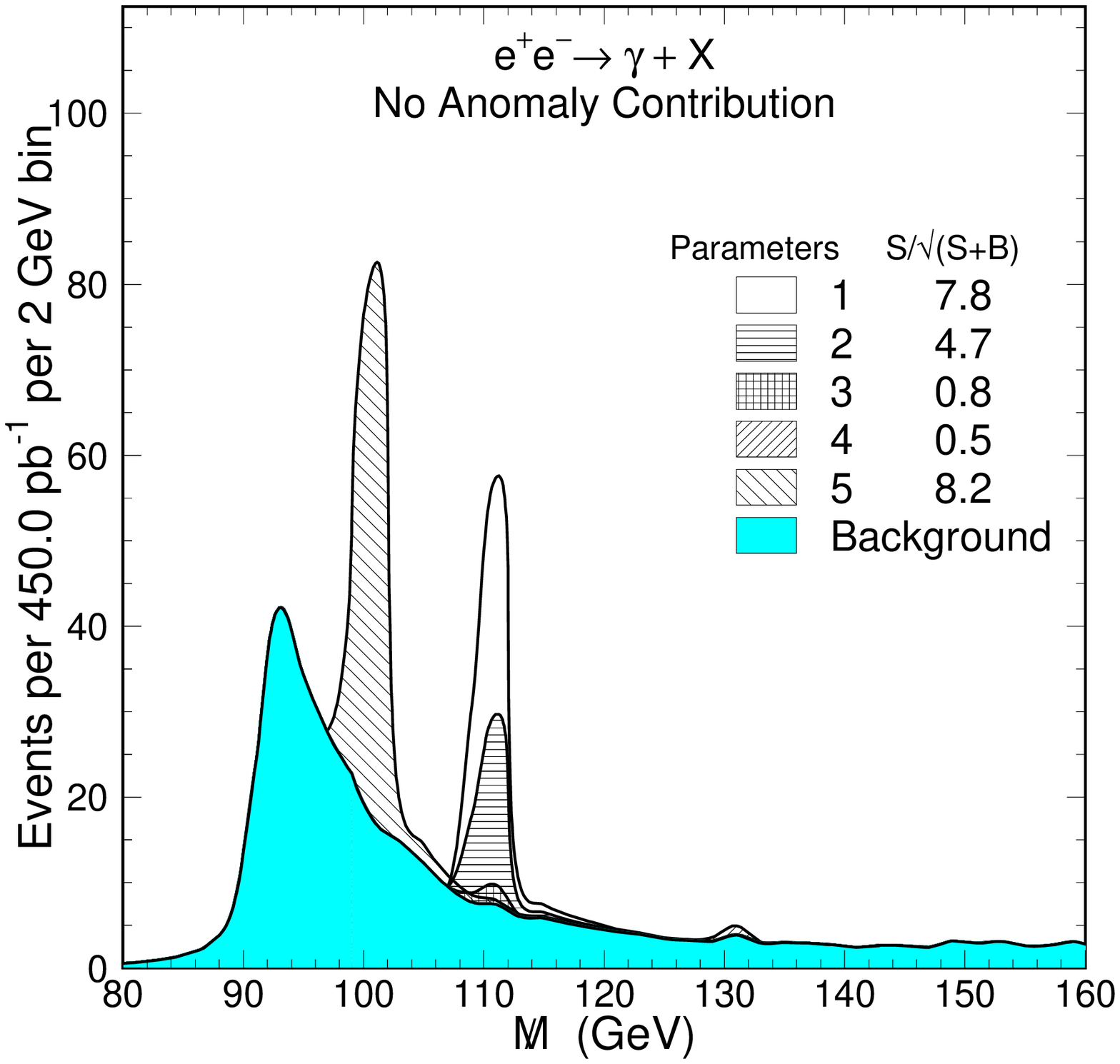}
\label{fig:pythia-b}}
\end{center}
\caption{The recoil mass spectrum in $\epem \to \photon\technipiboth$
at $\sqrt{s} = \unit[200]{GeV}$ as simulated at the particle level in
the TCSM using modifications to Pythia v6.202 as described in
Section~\ref{sec:pythia}.  In Figure~\ref{fig:pythia-a} we display the
background process, along with five different TCSM parameter sets: (1)
The baseline set with $M_V=M_A=\unit[200]{GeV}$, $\QU+\QD=5/3$,
$M_{\technipiboth}=\unit[110]{GeV}$, and
$M_{\technirho}=\unit[210]{GeV}$; (2) where $M_V=M_A=\unit[300]{GeV}$;
(3) where $\QU+\QD=0$; (4) where $M_{\technirho}=\unit[250]{GeV}$ and
$M_{\technipiboth}=\unit[130]{GeV}$; and (5) where
$M_{\technipiboth}=\unit[100]{GeV}$.  Figure~\ref{fig:pythia-b}
differs from Figure~\ref{fig:pythia-a} in the exclusion of the anomaly
coupling from the calculations.  The legends display the signal to
background significance ratios for each parameter set.}
\label{fig:pythia}
\end{figure}

We now discuss our Pythia studies of the process $\epem \to
\photon\technipiboth$ at the LEP collider.  The
kinematics of the process dictate that the photon is hard and more
central than would be expected in background processes.  We define
the signal to be a significant peak in the ``recoil mass'' recoiling
against the photon $\DS{M} = \sqrt{s - 2\sqrt{s}E_\photon}$ for
$E_\photon>\unit[10]{GeV}$ and $|\cos\theta_\photon|<0.7$. To reduce
backgrounds, the photon must pass an isolation requirement: there must
be no more than \unit[5]{GeV} of excess energy within an opening angle
of 30$^\circ$ centered on the photon.  Since the technipion is
expected to decay visibly, and predominantly to \qbottom\ quarks, we
will impose a \qbottom--tag to eliminate the potentially large
backgrounds from $\epem\to \photon\neutrino\antineutrino$.  We comment
later on how to generalize this search.

We simulated the signal at the particle level using Pythia v6.202
\cite{pythia}, with updates to the
technicolor simulation as specified in this paper.  The proposed
signature is a peak excess in the recoil mass distribution and a
loose \qbottom--tag in the rest of the event.  We are not sure how
stringent a \qbottom--tag needs to be imposed and have not included any
efficiency factors for the signal or fake rates from other quarks.  We
do not impose any kinematic cuts or jet--reconstruction algorithm on
the particles recoiling from the photon, but require only a displaced
vertex.

The only background included is from $\epem \to\photon \bbbar$.  To
account for the final--state radiation of photons off the
\qbottom--partons, the full 2-to-3 parton--level process is calculated
at the matrix element--level.  The parton level calculation is then
interfaced to Pythia, producing particle--level results that include
the effects of parton showering and hadronization.  After the
isolation cut on the photon, the results are in good agreement with
the standard Pythia simulation of $\epem \to
\photon+\photon^*/\Zstar$.  The implied suppression of radiation off
the \qbottom--quark arises from several effects: (1) the small charge
of the \qbottom, (2) the large \qbottom\ quark mass, which regulates
collinear emission, and (3) the kinematic constraints favoring a small
invariant mass of the ($\qbottom\photon$) or ($\antiqbottom\photon$)
systems, which is removed by the isolation cut.  Finally, we require
that at least one of the \qbottom--partons (after parton showering)
has a $p_T$ of at least \unit[5]{GeV}.  Assuming that displaced
vertices are detected with unit probability, this eliminates
backgrounds from light quarks.

The results are shown in Figure~\ref{fig:pythia-a}, assuming a collider
energy of \unit[200]{GeV} and $\unit[450]{pb^{-1}}$ of integrated
luminosity.  To demonstrate the variation with TCSM parameters, we
have chosen five parameter sets starting with the baseline (1):
$M_V=M_A=\unit[200]{GeV}$, $\QU+\QD=5/3$,
$M_{\technipiboth}=\unit[110]{GeV}$, and
$M_{\technirho}=\unit[210]{GeV}$.  The $\technipi$ and
$\technipiprime$ and, separately, the $\technirho$ and $\techniomega$
are assumed to be degenerate in mass.  The other parameter sets are
variations on this baseline, with all parameters as in (1) except: for
(2), $M_V=M_A=\unit[300]{GeV}$; for (3), $\QU+\QD=0$; for (4),
$M_{\technirho}=\unit[250]{GeV}$ and
$M_{\technipiboth}=\unit[130]{GeV}$; and, for (5),
$M_{\technipiboth}=\unit[100]{GeV}$.  The general TCSM condition $Q_U
- Q_D = 1$ obtains for all parameter sets.
Figure~\ref{fig:pythia-a} shows also
the significance defined as $S/\sqrt{S+B}$ for each of the parameter
sets.  

The baseline curve (1) includes a strong signal from the $\techniV$
poles just above the collider energy.  Comparison with set (2) shows
that the peak height scales as $M_V^{-2}$ as we would expect from
Equation~\ref{eqn:VTxsec}.  Comparison with set (4), confirms the
expected reduction of signal when the \techniV\ and \technipiboth\
masses are scaled up to put the collider energy well below the poles.
Set (3) allows us to infer that, as in figure \ref{fig:xsecs}, most of
the \techniV\ signal comes specifically from the \techniomega. Taking
$\QU+\QD=0$ decouples the \techniomega\ from the gauge bosons and
eliminates the \anomggpiboth\ coupling.  The branching fraction for
$\technirho \to \photon\technipiprime$ is also small for these TCSM
parameters, even though the \technirho\ is kinematically forbidden to
decay to a pair of \technipiboth; the dominant decay is to $\Wpart_L
\technipiboth$. Then the small signal in set (3) reflects the size of
contributions from the \anomgZpiboth\ anomaly and the $\technirho \to
\photon\technipi$ channel.  The only possible source of the much
larger peak in set (1) is the restored contribution of the
\techniomega.  Finally, in set (5), we deliberately open the (dominant
when present) decay channel $\technirho \to
\technipiboth\technipiboth$; this would normally be closed because of
large extended technicolor contributions to the \technipiboth\
masses~\cite{Lane:1999uh,Lane:1999uk}.  Although this significantly
decreases the $\technirho \to \photon\technipiboth$ branching ratio, a
strong signal of $\techniomega\to \photon\technipiboth$ still occurs
because \techniomega\ decays to two or three $\technipiboth$ remain
suppressed or forbidden.  Once again, the dominant role of the
\techniomega\ in the signal is confirmed.

Figure~\ref{fig:pythia-b} shows what the signals would look like if the
anomaly couplings \anomggpiboth\ and \anomgZpiboth\ were eliminated.
The peak heights and significances are clearly reduced in all cases.
For the parameter sets with the strongest signals (1,2,5), a
comparison with figure \ref{fig:pythia-a} confirms that the \techniomega\
resonance is largely responsible for making the \technipiboth\
production visible.  This is what we would expect from the results we
presented in Figure 2, because the mass of the resonance is just
slightly higher than the $\sqrt{s}$ of the collider.  Nonetheless, the
anomaly couplings make a contribution that can be large enough to
impact the limits extracted from the data.

Several further comments are in order.  First, the signature discussed
here is strongly dependent on the \techniomega\ properties, especially
the coupling $\photon \to \techniomega \to \photon\technipiboth$ which
is proportional to $\QU+\QD$.  In this respect, it is complementary to
the $\technirho \to \Wpm\technipiboth$ signatures
\cite{Eichten:1997yq}. On the other hand, the TCSM may be naive in its
assumption of certain mass degeneracies, and the \techniomega\ may be
significantly lighter than the \technirho, yielding the first
signature of technicolor.  Second, the proposed signature is not
inclusive, but assumes the $\technipiboth\to \bbbar$ branching ratio
is large (the observation of some visible energy is necessary to
remove the $\photon+\missingE$ background).  This is reasonable, but
the solution to the flavor problem may bring surprises in the
\technipi\ decay rates. There may also be a substantial rate for
$\technipiprime\to \gluon\gluon$ (this rate is already 30\% for
$\technipiprime$ with the default TCSM choices as encoded in Pythia),
and there even may be appreciable \technipi--\technipiprime mixing.
Therefore, while the first search should be tuned for the $\photon
\bbbar$ mode, we advocate a decay--independent search without the
\qbottom--tag. The background should be 4--5 times bigger, reducing
$S/B$ (which is a few--to--one for the $\photon \bbbar$ signature). It
will not not reduce the significance very much; we have defined it to
include a systematic error on the background.  For example, naively
scaling the background estimate by 5 would reduce $S/\sqrt{S+B}$ to
$6.6, 3.7, 0.4, 0.5$, and $6.2$ for parameter sets $(1)$--$(5)$,
respectively. This does not include the small increase in signal rate
from all decays of the technipions.

\section{Conclusions}
\label{sec:conclusions}

In the context of the TCSM, we have shown that both the anomaly and
kinetic mixing contributions should be included in analyses of
technipion production at lepton colliders.  We have provided analytic
formulas combining these contributions and used them to display the
predictions of the TCSM for a range of technihadron masses and
collider energies.  We have also performed Pythia simulations of
$\epem \to \photon\technipiboth$ including the modifications to the
TCSM described in this paper for five distinct sets of technihadron
masses and technifermion charges.  We find that resonant production of
technipions is necessary to ensure a visible signature at LEP II
energies for typical TCSM parameters, but that including the
non-resonant production will be important for setting accurate limits.
Finally, we note that measuring the recoil mass spectrum for $\epem
\to \photon\technipiboth$ production (which we found to proceed mainly
through the \techniomega) provides a technicolor search strategy that
is complementary to the $\technirho \to \Wpm\technipiboth$ channel.

\vspace{1in}
\begin{center}\textbf{Acknowledgments}\end{center}

\noindent We thank Guennadi Borisov and Francois Richard of DELPHI and
Markus Schumacher of OPAL for asking questions that led to the present
study and improvements in the TCSM.  We are also grateful to R. Sekhar
Chivukula and Markus Schumacher for comments on the manuscript.
K. Lane acknowledges the support of the Fermilab Theory Group through a
Frontier Fellowship.  This work was also supported in part by the
Department of Energy under grant DE-FG02-91ER40676 and by the National
Science Foundation under grant PHY-0074274.  FNAL is operated by
Universities Research Association Inc. under contract
DE-AC02-76CH03000.

\end{document}